\documentstyle[twoside,fleqn,espcrc2]{article}
\input{psfig}
\newcommand{\spur}[1]{\not\! #1 \,}

\newcommand{\AmS}{{\protect\the\textfont2
  A\kern-.1667em\lower.5ex\hbox{M}\kern-.125emS}}

\title{Beauty hadron lifetime ratios and the problem of $\tau(\Lambda_b)$}

\author{F. De Fazio\address{Istituto Nazionale di Fisica Nucleare, Sezione di 
        Bari,\\ 
        Via Amendola 173, I-70126 Bari, Italy}} 

\begin{document}

\begin{abstract}
We describe a theoretical approach to compute inclusive widths of heavy 
hadrons, based on a systematic expansion in the inverse powers of the heavy 
quark mass. The method reproduces quite well the experimental ratios
of $B$ mesons lifetimes. As for 
$\tau(\Lambda_b)/\tau(B_d)$, we present a QCD sum rule calculation of the 
${\cal O}(m_b^{-3})$ contribution  to $\tau(\Lambda_b)$; we conclude
that such correction is not responsible of the small experimental value of 
$\tau(\Lambda_b)/\tau(B_d)$.
\end{abstract}

\maketitle

\section{INCLUSIVE DECAYS OF HEAVY HADRONS}

A simple theoretical  way to consider 
 inclusive decays of hadrons containing one  heavy  quark 
  is the {\it Spectator Model}, 
which states that  only the heavy quark participates in the 
transition, while the light degrees of freedom 
are unaffected by the process. As a result, the model predicts
that  all the hadrons with the same heavy quark should have the 
same lifetime. 

On the other hand, the experimental ratios of beauty hadron lifetimes 
are \cite{blg}:

\begin{equation}
{\tau(B^-) \over \tau(B_d)}=1.06 \pm 0.04 \hskip 5pt
{\tau(B_s) \over \tau(B_d)}=0.99 \pm 0.05 \label{ratiomes} 
\end{equation}

\begin{equation}
{\tau(\Lambda_b) \over \tau(B_d)}=0.79 \pm 0.06 \; . \label{ratiolam}
\end{equation}

\noindent Eq.(\ref{ratiolam}) shows  that the ratio 
$\tau(\Lambda_b) / \tau(B_d)$ significantly differs from the 
Spectator Model prediction.

One can employ a more refined theoretical approach, based on QCD, 
to compute inclusive decay widths of heavy hadrons; the method consists 
in an 
expansion in powers of $m_Q^{-1}$ \cite{misha}. 

By using the optical theorem, the 
inclusive width of a hadron containing the heavy 
quark $Q$ into a final state with 
assigned quantum numbers $f$ can be written as:

\begin{equation}
\Gamma(H_Q \to X_f)={2\; Im<H_Q |{\hat T}|H_Q> \over 2 M_{H_Q}} \label{incl}
\end{equation}

\noindent 
where ${\hat T}(Q \to X_f \to Q)$ is the transition operator describing 
$Q$ with the same momentum in the 
initial and final state:

\begin{equation}
{\hat T}=i \; \int d^4x \; T[{\cal L}_w(x){\cal L}_w^{\dag}(0)] \hskip 3 pt . 
\label{trans}
\end{equation}

\noindent ${\cal L}_w$ is the effective weak lagrangian for
the decay $Q \to X_f$. 
The time ordered product in
(\ref{trans}) is computed by an operator product expansion (OPE):

\begin{equation}
{\hat T}=\sum_i C_i {\cal O}_i \label{ope}
\end{equation}

\noindent where ${\cal O}_i$ are local operators ordered 
by increasing dimension, and 
$C_i$ are coefficients ordered by increasing powers of $1/m_Q$.
As a result, the inclusive width of a heavy hadron $H_b$ (from 
now on we will refer to the beauty sector) is represented by a sum of 
expectation values over $H_b$ of operators with increasing dimension:

\begin{eqnarray} 
\Gamma(H_b \to X_f)={G_F^2 m_b^5 \over 192 \pi^3}|KM|^2 \cdot \hskip 1.3 cm && 
\nonumber \\
\Big[c_3^f {<H_b|{\bar b}b|H_b> \over 2 M_{H_b}}  \hskip 3.2 cm && \nonumber \\
+{c_5^f \over m_b^2} 
{<H_b|{\bar b} i g_s \sigma \cdot G b|H_b> \over 2 M_{H_b}} 
\hskip1.7cm && \nonumber \\
+\sum_i {c_6^{f(i)} \over m_b^3} 
{<H_b|{\cal O}_i^6|H_b> \over 2 M_{H_b}} +{\cal O}\Big({1 \over m_b^4} \Big) 
\Big]  .&& \label{ris}
\end{eqnarray}
 
\noindent
$KM$ is the CKM matrix element relevant in the considered decay. 

The first operator in (\ref{ris}) is ${\bar b}b$, which has dimension 
$D=3$; next, the chromomagnetic operator ${\cal O}_G={\bar b}{g \over 2} 
\sigma_{\mu \nu} G^{\mu \nu}b$, responsible of the mass splitting between 
hadrons which differ only for the heavy quark spin orientation, has $D=5$; 
finally, $O_i^6$ have $D=6$ and contribute to ${\cal O}(m_b^{-3})$ 
in the expansion.
$<{\bar b}b>_{H_b}=<H_b|{\bar b} b |H_b>/2 M_{H_b}$ 
can be further reduced by using the heavy quark 
equation of motion  in the limit $m_b \to \infty$:

\begin{eqnarray}
<{\bar b}b>_{H_b}&=& 1+{<{\cal O}_G>_{H_b} \over 2 m_b^2} \nonumber \\
&-&{<{\cal O}_\pi>_{H_b} \over 2 m_b^2} +{\cal O}\Big({1 \over m_b^3}\Big) .
\label{bbarb}
\end{eqnarray}

\noindent ${\cal O}_\pi={\bar b}(i {\vec D})^2b$ is the heavy quark 
kinetic energy operator due to its residual motion inside the hadron.

Some important features should be noticed in (\ref{ris}). The first term of 
the expansion reproduces the Spectator Model result; ${\cal O}(m_b^{-1})$ 
terms  are absent \cite{cgg,buv} 
since $D=4$ operators cannot  contribute, being reducible, by 
equation of motion, to the $D=3$ operator ${\bar b}b$; finally, ${\cal 
O}_G$ and ${\cal O}_\pi$ are spectator blind, i.e. they are not sensitive 
to the light flavour.

${\cal O}(m_b^{-3})$ terms come from four quark operators, which can be 
classified as follows \cite{neub}:

\begin{eqnarray}
O^q_{V-A}&=&{\bar b}_L \gamma_\mu q_L {\bar q}_L \gamma_\mu b_L \nonumber \\
O^q_{S-P}&=&{\bar b}_R q_L {\bar q}_L b_R \nonumber \\
T^q_{V-A}&=&{\bar b}_L \gamma_\mu t^a q_L {\bar q}_L \gamma_\mu t^a b_L 
\nonumber \\
T^q_{S-P}&=&{\bar b}_R t^a q_L {\bar q}_L t^a b_R \label{t} 
\end{eqnarray}
 
\noindent where $t^a=\lambda^a/2$, $\lambda^a$ being the Gell-Mann matrices.
Their contribution accounts for the presence of the spectator 
quark, which participates in the decay through the mechanisms known as weak 
annihilation and Pauli interference (the latter depends on the presence of two 
identical quarks in the final state).

In order to use the result (\ref{ris}) one has to know the matrix elements of 
all the previously mentioned operators. Let us define:

\begin{equation}
\mu_G^2(H_b)={<H_b|{\cal O}_G|H_b> \over 2 M_{H_b}} \;, \label{mug}
\end{equation}

\begin{equation}
\mu_\pi^2(H_b)={<H_b|{\cal O}_\pi|H_b> \over 2 M_{H_b}} \; . \label{mupi}
\end{equation}

\noindent $\mu^2_G$ depends on the spin $J$ of $H_b$: 
$\mu_G^2=-2[J(J+1)-{3 \over 2}] \lambda_2$. It enters with
$\mu_\pi^2$ in the mass formula for a heavy hadron in the $m_b \to \infty$ 
limit:

\begin{equation}
M_{H_b}=m_b+{\bar \Lambda}+{\mu_\pi^2 -\mu_G^2 \over 2 m_b}
+{\cal O}\Big({1 \over m_b^2}\Big)   \; . \label{mass}
\end{equation}

\noindent The parameters ${\bar \Lambda}$, $\mu_\pi^2$, $\lambda_2$ are 
independent of the heavy quark mass.
For $B$ mesons, $\mu_G^2$ can be derived experimentally,
since it is related to the mass splitting: $\mu_G^2(B)=3(M_{B^*}^2-M_B^2)/4 
\simeq 0.36 \;\;GeV^2$. On the other hand: $\mu_G^2(\Lambda_b)=0$, 
since the light degrees of freedom have zero total angular 
momentum relative to the heavy quark.

Various theoretical determinations exist for $\mu_\pi^2(B_d)$. 
From the mass formula (\ref{mass}) and using experimental data on the 
mass of the $\Lambda_b$ baryon, one obtains:

\begin{equation}
\mu_\pi^2(B_d)-\mu_\pi^2(\Lambda_b)=0.002 \pm 0.024 \; GeV^2 \label{diffmu}
\end{equation}

\noindent and hence it can be assumed: $\mu_\pi^2(B_d)\simeq 
\mu^2_\pi(\Lambda_b)$\footnote{This is confirmed by  QCD 
sum rule  estimates \cite{braun,loro}.}.

In order to evaluate the matrix elements of $D=6$ 
operators, different approaches must be used for 
mesons and baryons. For $B$ mesons, factorization 
approximation gives:

\begin{eqnarray}
{<B_q|O^q_{V-A}|B_q> \over 2 M_{B_q} }=\hskip 3.3cm&&\nonumber \\
{<B_q|O^q_{S-P}|B_q> \over 2 M_{B_q} } 
\Big({m_b+m_q \over M_{B_q}} \Big)=f^2_{B_q} {M_{B_q} \over 8}&& \nonumber \\
<B_q|T^q_{V-A}|B_q>= <B_q|T^q_{S-P}|B_q> =0 \;\;\;&& \label{fact} 
\end{eqnarray}

\noindent $f_{B_q}$ being the $B_q$ decay constant. 
The same technique cannot be applied 
to baryons, so that one must use other approaches. 
In the case of $\Lambda_b$, heavy quark symmetries reduce the 
relevant operators to two \cite{neub}:

\begin{eqnarray}
{\tilde O}^q_{V-A}&=&{\bar b}_L \gamma^\mu b_L {\bar q}_L \gamma_\mu q_L
\nonumber \\
O^q_{V-A}&=&{\bar b}_L \gamma^\mu q_L {\bar q}_L \gamma_\mu b_L
\; . \label{fourq} 
\end{eqnarray}

\noindent One can parameterize the matrix elements as follows:

\begin{eqnarray}
{<{\tilde O}_{V-A}^q>_{\Lambda_b} \over 2 M_{\Lambda_b}}&=&
{f^2_B M_B \over 48}r 
\nonumber \\
{<O_{V-A}^q>_{\Lambda_b} \over 2 M_{\Lambda_b}}&=&-{\tilde B}
{<{\tilde O}_{V-A}^q>_{\Lambda_b} \over 2 M_{\Lambda_b}} \; ;\label{otilde}
\end{eqnarray}

\noindent in the valence quark approximation: ${\tilde B}=1$.

It should be stressed that only large values of the parameter $r$ in 
(\ref{otilde}) $(r=3-4)$ could explain the observed difference between 
$\tau(\Lambda_b)$ and $\tau(B_d)$. Calculations based on quark model 
\cite{guberina} or exploiting the experimental splitting: 
$M_{\Sigma_b^*}-M_{\Sigma_b}$ \cite{rosner} 
give values of $r$ smaller than required. 
In order to have a different theoretical estimate, we 
computed by QCD sum rules the matrix elements of the relevant $D=6$ operators 
over $\Lambda_b$. This calculation will be described in the next section.

\section{QCD SUM RULE CALCULATION OF $<{\tilde O}^q_{V-A}>_{\Lambda_b}$}

The starting point to evaluate $<{\tilde O}^q_{V-A}>_{\Lambda_b}$ by QCD sum 
rules within the Heavy Quark Effective Theory (HQET) is the 
correlation function:

\begin{eqnarray}
\Pi_{CD}(\omega,\omega^\prime)=(1+{\spur v})_{CD} \Pi (\omega,\omega^\prime) 
&&\nonumber \\
=i^2 \int dx \; dy \;\;e^{i \omega (v \cdot x)-i \omega^\prime (v \cdot y)} 
\;\;\;\;\;\;&&
\nonumber  \\
<0|T[J_C(x) {\tilde O}_{V-A}^q(0) J_D(y)]|0> &&   \label{corr}
\end{eqnarray}

\noindent where $C$, $D$ are Dirac indices; the variable $\omega$ 
$(\omega^\prime)$ is related to the residual momentum of the incoming 
(outgoing) baryonic current $p^\mu=m_b v^\mu+k^\mu$ with $k^\mu=\omega v^\mu$.
A suitable interpolating field for $\Lambda_b$, in the  
 limit $m_b \to \infty$, is given by \cite{shur}:

\begin{equation}
J_C(x)=\epsilon^{ijk}(q^{Ti}(x)\Gamma\tau q^j(x))(h^k_v)_C(x) 
\label{corrente}
\end{equation}

\noindent where $T$ means the transpose, $i$, $j$, $k$ are colour indices, 
$h_v$  is  the effective heavy quark field, $\Gamma=C \gamma_5 (1+ b{\spur v})$ 
and $C$ is the 
charge conjugation operator.  $\tau$ is the $\Lambda_b$ light flavour 
matrix:

\begin{equation}
\tau= \frac{1}{\sqrt 2}
\left( \begin{array}{cc} 0 & 1 \\ -1 & 0 \end{array} \right)\;.
\end{equation}

\noindent The parameter $b$ is fixed to $b=1$ from the QCD sum rules 
analysis of $f_{\Lambda_b}$ \cite{loro}, defined by: $<0|J_C|\Lambda_b(v)>
=f_{\Lambda_b} (\psi_v)_C $ ($\psi_v$ is the spinor for a $\Lambda_b$ of 
four-velocity $v$).

The hadronic representation of the correlator (\ref{corr}) can be obtained 
saturating it by baryonic states, and considering the double pole contribution 
in the variables $\omega$ and  $\omega^\prime$:

\begin{eqnarray}
\Pi^{had}(\omega, \omega^\prime) = 
\langle{\tilde {\cal O}^q_{V-A}}\rangle_{\Lambda_b}
{f^2_{\Lambda_b} \over 2} \times \nonumber \\
{1 \over (\Delta_{\Lambda_b} - \omega)(\Delta_{\Lambda_b} - \omega^\prime)} 
+ \dots \hskip 0.6 cm \label{phad}
\end{eqnarray}

\noindent at the value  $\omega=\omega^\prime=\Delta_{\Lambda_b}$. 
$\Delta_{\Lambda_b}$ represents the binding energy of 
the light degrees 
of freedom in $\Lambda_b$: 
$M_{\Lambda_b}= m_b + \Delta_{\Lambda_b}$
and must be derived within the same QCD sum rule framework.
On the other hand, in the Euclidean region, for negative values of $\omega$, 
$\omega^\prime$, the correlator (\ref{corr}) can be computed in QCD, 
in terms of a perturbative contribution and of vacuum condensates. The result 
can be written in a dispersive form:

\begin{equation}
\Pi^{OPE}(\omega, \omega^\prime)= \int d\sigma d\sigma^\prime 
{\rho_\Pi(\sigma, \sigma^\prime) \over(  \sigma - \omega ) (\sigma^\prime
- \omega^\prime)} \label{pope}
\end{equation}

\noindent where possible subtraction terms have been omitted. The explicit 
expression of the computed spectral function $\rho_\Pi$ can be found in 
\cite{noi}.

A sum rule for $<{\tilde O}^q_{V-A}>$ can be derived by equating $\Pi^{had}$ 
and $\Pi^{OPE}$.
Moreover, invoking a global duality ansatz,
the contribution of the higher resonances and of the continuum in 
$\Pi^{had}$ in (\ref{phad}) can be  modeled as the QCD term
outside the region 
$0 \le \omega \le \omega_c$,
$0 \le \omega^\prime \le \omega_c$, with $\omega_c$ an effective threshold.
Finally, the application of a double Borel transform  
to both $\Pi^{OPE}$ and $\Pi^{had}$  in the momenta $\omega, \omega^\prime$:

\begin{eqnarray} 
{\cal B}(E_1)\frac{1}{\sigma-\omega}&=&\frac{1}{E_1} \; e^{-\sigma/E_1},
\nonumber \\
{\cal B}(E_1)\frac{1}{\Delta-\omega}&=&\frac{1}{E_1} \; e^{-\Delta/E_1}, 
\label{borel}
\end{eqnarray}

\noindent 
(and similar for $\omega^\prime$ with the Borel parameter $E_2$) 
allows us to remove the subtraction terms 
in (\ref{pope}), that are polynomials in $\omega$
or $\omega^\prime$ (the Borel transform of a polynomial vanishes). 
Moreover,
the convergence of the OPE is factorially improved by the transform, and
the contribution of the low-lying resonances in $\Pi^{had}$ is enhanced for
low values of the Borel variables.
The symmetry of the spectral 
function in $\sigma$, $\sigma^\prime$ suggests the choice $E_1=E_2=2 E$.
The final sum rule reads:

\begin{eqnarray}
{f^2_{\Lambda_b} \over 2} (1 + b)^2 \exp(- {\Delta_{\Lambda_b} \over E}) 
\langle \tilde {\cal O}^q_{V-A}\rangle_{\Lambda_b} \hskip 1.4 cm
&&\nonumber \\ =
\int_0^{\omega_c} \int_0^{\omega_c} d \sigma d \sigma^\prime
\exp(-{ \sigma + \sigma^\prime \over 2 E} ) 
\rho_\Pi(\sigma, \sigma^\prime)
\;\;\; .&&\label{sr}
\end{eqnarray}

\begin{figure}
\mbox{\psfig{file=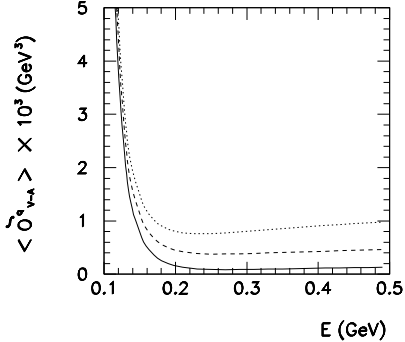,height=1.8in}}
\vskip -0.3cm
\caption{Sum rule for $<{\tilde O}_{V-A}^q>_{\Lambda_b}$ as a function 
of the Borel variable $E$. \label{th271_fig1}}
\end{figure}

The result of the numerical analysis of (\ref{sr}) is depicted 
in Fig. \ref{th271_fig1}.
A stability window is observed, starting at 
$E\simeq 0.2 \; GeV$ and continuing towards large values of $E$;
in this range the result for 
$\langle \tilde {\cal O}^q_{V-A}\rangle_{\Lambda_b}$ is independent of the 
external parameter $E$.
The variation of the result with $E$ and with 
$\omega_c$ provides us with an estimate of 
the accuracy of the numerical outcome. 
We find \cite{noi}:
\begin{equation}
\langle\tilde {\cal O}^q_{V-A}\rangle_{\Lambda_b} \simeq (0.4 - 1.20) \times 
10^{-3} \; GeV^3 \;\;\; , \label{res}
\end{equation}
a result corresponding to  $r\simeq 0.1~-~0.3$.

As for $\tilde B$ in (\ref{otilde}),
since our 
computational scheme considers only valence quark processes,
 a sum rule for the matrix element in (\ref{otilde}) would give
$\tilde B=1$. The  explicit calculation 
confirms this conclusion. 
Using the computed $r$ and ${\tilde B}$ in the formulae in ref. \cite{neub}, we 
get: $\tau(\Lambda_b)/\tau(B_d)\ge 0.94$.

\section{CONCLUSIONS}

We presented a method  based on the $1/ m_Q$ expansion to compute 
inclusive widths of heavy hadrons.  We also described the 
 QCD sum rule calculation 
of the matrix element $<{\tilde O}^q_{V-A}>_{\Lambda_b}$ 
contributing to ${\cal O}(m_b^{-3})$ to the $\Lambda_b$ lifetime. 
The result: $\tau(\Lambda_b)/\tau(B_d)\ge 0.94$ implies that such 
correction  cannot explain the observed difference
between $\tau(\Lambda_b)$ and $\tau(B_d)$. If the experimental data
 will be confirmed, a theoretical reanalysis of 
the problem would be required, as suggested for example in 
\cite{altarelli,dom}.

\vskip 0.3 cm
I thank P. Colangelo for his collaboration. Useful discussions 
with G. Nardulli are also acknowledged.

\end{document}